\documentclass[sigconf]{acmart}

\AtBeginDocument{%
  \providecommand\BibTeX{{%
    \normalfont B\kern-0.5em{\scshape i\kern-0.25em b}\kern-0.8em\TeX}}}

\usepackage{multirow}
\usepackage{xcolor}
\usepackage{svg}
\usepackage{url}
\usepackage{subfigure}

\copyrightyear{2024}
\acmYear{2024}
\setcopyright{rightsretained}
\acmConference[UbiComp Companion '24]{Companion of the 2024 ACM International Joint Conference on Pervasive and Ubiquitous Computing }{October 5--9, 2024}{Melbourne, VIC, Australia}
\acmBooktitle{Companion of the 2024 ACM International Joint Conference on Pervasive and Ubiquitous Computing (UbiComp Companion '24), October 5--9, 2024, Melbourne, VIC, Australia}
\acmDOI{10.1145/3675094.3678423}
\acmISBN{979-8-4007-1058-2/24/10}

\begin{document}


\title[Leveraging Large Language Models for Generating Mobile Sensing Strategies \\ in Human Behavior Modeling]{Leveraging Large Language Models for Generating Mobile
Sensing Strategies in Human Behavior Modeling}

\author{Nan Gao}
\orcid{0000-0002-9694-2689}
\affiliation{%
  \institution{Tsinghua University}
  \city{Beijing}
  \country{China}
}
\affiliation{%
  \institution{University of New South Wales (UNSW)}
  \city{Sydney}
  \country{Australia}
  \postcode{1466}
}
\orcid{0000-0002-9694-2689}
\email{nangao@tsinghua.edu.cn}

\author{Zhuolei Yu}
\affiliation{%
  \institution{Tsinghua University}
  \city{Beijing}
  \country{China}
}
\orcid{0009-0003-9132-9759}
\email{yuzl21@mails.tsinghua.edu.cn}

\author{Yue Xu}
\affiliation{%
  \institution{Tsinghua University}
  \city{Beijing}
  \country{China}
}
\orcid{0009-0003-8440-5549}
\email{yue-xu22@mails.tsinghua.edu.cn}

\author{Chun Yu}
\authornote{Corresponding author.}
\affiliation{%
  \institution{Tsinghua University}
  \city{Beijing}
  \country{China}
}
\orcid{0000-0003-2591-7993}
\email{chunyu@mail.tsinghua.edu.cn}

\author{Yuntao Wang}
\affiliation{%
  \institution{Tsinghua University}
  \city{Beijing}
  \country{China}
}
\orcid{0000-0002-4249-8893}
\email{yuntaowang@tsinghua.edu.cn}

\author{Flora D. Salim}
\orcid{0000-0002-1237-1664}
\affiliation{%
  \institution{University of New South Wales (UNSW)}
  \city{Sydney}
  \country{Australia}
  \postcode{1466}
}
\orcid{0000-0002-4249-8893}
\email{flora.salim@unsw.edu.au}

\author{Yuanchun Shi}
\affiliation{%
  \institution{Tsinghua University}
  \city{Beijing}
  \country{China}
}
\orcid{0000-0003-2273-6927}
\email{ shiyc@tsinghua.edu.cn}
\renewcommand{\shortauthors}{Nan Gao et al.}


\begin{abstract}

Mobile sensing plays a crucial role in generating digital traces to understand human daily lives. However, studying behaviours like mood or sleep quality in smartphone users requires carefully designed mobile sensing strategies such as sensor selection and feature construction. This process is time-consuming, burdensome, and requires expertise in multiple domains. Furthermore, the resulting sensing framework lacks generalizability, making it difficult to apply to different scenarios. In the research, we propose an automated mobile sensing strategy for human behaviour understanding. First, we establish a knowledge base and consolidate rules for data collection and effective feature construction. Then, we introduce the multi-granular human behaviour representation and design procedures for leveraging large language models to generate strategies. Our approach is validated through blind comparative studies and usability evaluation.  Ultimately, our approach holds the potential to revolutionise the field of mobile sensing and its applications.

\end{abstract}

\ccsdesc[500]{Applied computing}
\keywords{Self-report survey, Mobile sensing, Human behavioural modelling, Large language models, Human-computer collaboration}


\maketitle



\section{Introduction}

The development of the Internet of Things (IoT) has transformed how we capture and analyze digital traces of daily life. Mobile sensing, a form of passive sensing using smartphone sensor data, plays a crucial role in this transformation. By leveraging data from software and hardware sensors, mobile sensing provides a comprehensive understanding of human behaviours \cite{wampfler2022affective, wang2020social, gao2019predicting}. Compared to wearable and environmental sensing, mobile sensing offers unobtrusive, long-term data collection in real-world settings, reducing user burden and providing convenience without additional devices \cite{laport2020review}. Additionally, multiple sensors in mobile devices yield rich, diverse data, facilitating a contextual understanding of the surroundings.

Recently, mobile sensing has become popular for understanding human behaviours, such as affective states \cite{wampfler2022affective}, academic performance \cite{wang2015smartgpa}, life satisfaction \cite{yuruten2014predictors}, and personality \cite{gao2019predicting}. It serves as an effective \textit{Quantified-Self} tool \cite{lee2014s} to enhance self-awareness and well-being, with applications in health monitoring and personalized services. For example, Gao et al. \cite{gao2019predicting} predicted Big-5 personality traits using call logs, message logs, and accelerometer data. Wampfler et al. \cite{wampfler2022affective} predicted affective states using touch and IMU data. Wang et al. \cite{wang2015smartgpa} used activity, conversational interaction, and mobility data to predict college students' GPA.

However, understanding human behaviours through mobile sensing presents significant challenges, especially for complex behaviours like well-being and personality traits. One one hand, researchers need a deep understanding of relevant domain knowledge (e.g., psychology \cite{gao2019predicting, hong2022depressive}, medicine \cite{obuchi2020predicting}, education \cite{wang2015smartgpa,gao2020n}) to extract pertinent features effectively. Expertise in sensor combinations, battery optimization, device settings, and data-driven modeling is essential for accurate models. For instance, a study on social functioning in individuals with schizophrenia \cite{wang2020social} used various mobile sensing data types and extracted features related to social functioning, highlighting the need for domain knowledge and machine learning skills.

On the other hand, traditional mobile sensing studies often focus on specific research objectives (e.g., measuring depression during COVID-19 \cite{nepal2022covid}, identifying time-killing moments on smartphones \cite{chen2023you}, predicting weekend nightlife drinking behaviour \cite{meegahapola2021examining}), resulting in frameworks that lack generalizability. This makes it challenging to apply them to different scenarios and participants, especially with minor variations in sensor usage.

Therefore, we aim to explore the automation of mobile sensing strategies for dynamic research objectives. While automation has been implemented in traditional modeling tasks (e.g., AutoML \cite{hutter2019automated} and Auto-Sklearn \cite{feurer2020auto}), most focus on traditional tabular data rather than mobile sensing settings and do not effectively utilize semantic information. Our research questions are: \textit{
1. What specific types of data should be collected to achieve different research objectives using mobile sensing technologies?
2. How can the collected data be effectively utilized to generate meaningful features that align with the research objective?
3. Which models can be utilized, and what is the estimated performance based on the research objective?}

To address these questions, we propose an automated mobile sensing strategy generation system. We reviewed mobile sensing studies from top venues, building a knowledge base. From this, we consolidated rules for feature construction, sensor selection, and model suggestions. We also developed a multi-granular human behavior decomposition mechanism to understand behaviors at varying levels. Large Language Models (LLMs) were utilized in five steps of strategy generation. The system outputs automated mobile sensing strategies that dynamically respond to user inquiries. Our contributions are as follows:


\begin{itemize}

    \item We establish a mobile sensing knowledge base from 55 studies in reputable venues such as CHI and IMWUT, identifying rules for effective feature construction and sensor selection. 
    
    \item We develop a multi-granular human behaviour representation mechanism for understanding behaviours in mobile sensing settings, aiding in effective feature construction.
    
    \item We propose an automated mobile sensing strategy that provides suggestions for data selection, feature construction, model building and performance estimation. 
    
\end{itemize}



\section{Related works}
\label{sec: relatedworks}

\subsection{Modelling Human Behaviours using Mobile Sensing Technologies}
\label{subsec: human behaviours}


Mobile sensing technologies revolutionize understanding human behavior, enabling predictions of personality traits \cite{gao2019predicting}, depression \cite{wang2022first}, stress-resilience \cite{adler2021identifying}, social anxiety \cite{rashid2020predicting}, and schizophrenia \cite{wang2017predicting}. They also explore links with alcohol consumption \cite{meegahapola2021examining}, behavior post-promotion \cite{nepal2020detecting}, time-killing on smartphones \cite{chen2023you}, and notification response time \cite{heinisch2022investigating}. Capturing real-time data in natural settings, mobile sensing offers unprecedented insights into human life.


While mobile sensing infers various aspects of human behavior, each requires comprehensive study design, data collection, and feature construction. Researchers typically invest significant time in these areas.
Traditionally, data is collected via background apps (e.g., SensingKit \cite{katevas2014poster}, AWARE \cite{ferreira2015aware}, AWARE-Light \cite{van2023awarelight}, CARP \cite{bardram2020carp}), but excessive data collection poses challenges like unused data, battery drain, and privacy concerns, reducing participant willingness and requiring extensive post-processing \cite{van2023awarelight}. Limited data collection, however, restricts understanding due to budget and ethical constraints. We propose \textbf{RQ1} to optimize data collection.

Feature engineering, creating new features from raw data \cite{kuhn2019feature}, is time-consuming and requires multi-domain expertise. Effective features enhance model performance, while poor features yield poor results. Traditional features, like statistical measures \cite{dehkordi2020feature}, have limited effectiveness due to human behavior's complexity. We propose \textbf{RQ2} to enable automated feature construction, reducing reliance on human expertise and streamlining data collection by rationalizing sensor data selection.


Accurate prediction models are essential for understanding behavior, approached through regression \cite{gao2019predicting, wang2020social,gao2020n} or classification \cite{chen2023you,meegahapola2023generalization}. Traditional models include \textit{Random Forest} (RF) \cite{segal2004machine}, \textit{Gradient Boosting} (GB) \cite{friedman2002stochastic}, and \textit{Naive Bayes} (NB) \cite{rish2001empirical}. Neural networks and deep learning are limited by small participant samples. Researchers need to estimate model performance before studies. We propose \textbf{RQ3} to help identify suitable models, understand expected performance, aid informed decisions, and recognize prediction limitations.

\subsection{AutoML and Large Language Models}

Auto Machine Learning (AutoML) \cite{hutter2019automated} offers automated solutions for identifying efficient machine learning pipelines, with notable successes including AutoSklearn \cite{feurer2020auto}, Auto-WEKA \cite{kotthoff2019auto}, and Auto-Pytorch \cite{imambi2021pytorch}. However, these focus on traditional features, overlooking semantic information. 
Large Language Models (LLMs) excel in natural language processing \cite{tornede2023automl}, encapsulating a wealth of domain knowledge. 
Hollmann \cite{hollmann2023gpt} proposed CAAFE, leveraging LLMs for semantically meaningful feature engineering from dataset descriptions, but focused on simple tabular data. Applying such methods to complex sensing data and human behavior research remains underexplored.
The prevalence of LLMs has opened up new possibilities for understanding human needs by exploring behavior-related variable correlations. Their embedded domain knowledge can automate data science tasks involving intricate contextual information. This intersection of AutoML and LLMs presents a promising direction for future research.


\section{Methodology}
\label{sec: base}

\subsection{Construction of Knowledge Base}

We focused on papers using only mobile sensing, excluding other sources like wearables, to construct our knowledge base. We selected articles from top venues in mobile sensing and ubiquitous computing: CHI (Conference on Human Factors in Computing Systems) and IMWUT (Proceedings of the ACM on Interactive, Mobile, Wearable, and Ubiquitous Technologies). The selection process included: 1. Accessing the ACM advanced search website \footnote{\url{https://dl.acm.org/search/advanced}}. 
2.	Search Within: ‘Title’ = (mobile OR smartphone) AND (sensing OR sensors OR sensor OR sense) NOT (wearable OR wristband OR desktop OR wrist-worn OR environmental OR environment OR laptop)
3.	Apply the filters successively to ensure the inclusion of relevant research articles: 
a)	Select ‘UbiComp: Ubiquitous Computing’ AND ‘Research Article’.
b)	Select ‘CHI: Conference On Human Factors In Computing Systems’ AND ‘Research Article’.
c)	Select ‘Proceedings Of The ACM On Interactive, Mobile, Wearable And Ubiquitous Technologies’ AND ‘Research Article’
In total, we collected 121 papers: 42 from IMWUT, 22 from CHI, and 57 from Ubicomp. After meticulous review, we retained 55 papers that exclusively used mobile sensing and focused on human behaviour.

\begin{table}
\renewcommand{\arraystretch}{0.9}
\footnotesize
\caption{An overview of features components summarised from the knowledge base}
\label{tab:features}
\setlength{\tabcolsep}{1pt}
\begin{tabular}{llll}
\toprule
\textbf{Component}                  & \textbf{Category}             & \textbf{Descriptions}   & \textbf{Example values}                                                                                                  \\ \midrule
\multirow{3}{*}{\textit{Time span}} & \multirow{2}{*}{Duration}     & Daily epoches           & Morning, afternoon, night                                                                                                \\
                                    &                               & Past to present         & In the last 30 minutes                                                                                                   \\ \cmidrule{2-4}
                                    & Periodicity                   & Recurrence              & Daily, weekly, monthly                                                                                                   \\ \midrule
\multirow{14}{*}{\textit{Metrics}}  & \multirow{5}{*}{Statistical}  & Central tendency        & Mean, median, mode                                                                                                       \\
                                    &                               & Dispersion              & Standard deviation, variance, range                                                                                      \\
                                    &                               & Shape                   & Skewness, kurtosis                                                                                                       \\
                                    &                               & Direct                  & Temperature, screen on state, location                                                                                   \\\cmidrule{3-4}
                                    &                               & Others                  & \begin{tabular}[c]{@{}l@{}}Count, magnitude, sum, slope, max, min \\ frequency, ratio, proportion\end{tabular}                    \\ \cmidrule{2-4}
                                    & \multirow{2}{*}{Regularity}   & Regularity              & \begin{tabular}[c]{@{}l@{}}Mean Squares Successive Difference (MSSD),\\ regularity index, consistency score\end{tabular} \\ \cmidrule{3-4}
                                    &                               & Circadian rhythms       & Same as above                                                                                                            \\ \cmidrule{2-4}
                                    & \multirow{2}{*}{Relation} & Correlation             & Pearson, Spearman, Kendall tau correlation                                                                               \\
                                    &                               & Ranking                 & The most frequent place visited                                                                                          \\ \cmidrule{2-4}
                                    & Diversity                     & Diversity of values & Shannon entropy                                                                                                          \\\cmidrule{2-4}
                                    & Similarity                    & Similarity              & Cosine, Jaccard, Hamming distance                                                                                       \\\cmidrule{2-4}
                                    & Spatial                       & Spatial                 & Distance, density, location                                                                                              \\\cmidrule{2-4}
                                    
                                    & Temporal                      & Temporal                & Duration, frequency, trend                                                                                              \\ \cmidrule{2-4}
                                    & Other                         & Other measures          & \begin{tabular}[c]{@{}l@{}}Fast Fourier Transform (FFT), \\ Mel Frequency Cepstral Coefficient (MFCC)\end{tabular}       \\ \bottomrule 
\end{tabular}
\end{table}

\subsection{Overview of Data Sources}
\label{subsec: data source}

From the reviewed papers, we identified commonly used sensors for mobile sensing studies, excluding those used in fewer than two papers due to potential data collection difficulties. We also standardized the names of data sources for consistency. The most commonly used sensors are:

\begin{itemize}
    \item \textbf{Hardware Sensors}. Integral components in mobile devices that monitor physical activities. Common sensors include: \textit{[Accelerometer, Gyroscope, Light, Magnetometer, Gravity, Temperature, Humidity, Orientation, Barometer, Proximity, Microphone, Bluetooth, WiFi]}.
    \item \textbf{Software Senors}. Derived from hardware sensors combined with software models to deduce new variables. Common sensors include: \textit{[Application, Calls, Message, GPS/Location, Notification, Keyboard]}.
    \item \textbf{Contextual Information}. Provides insight into the surrounding environment or circumstances of device usage, essential for understanding user behaviour and preferences. Common data includes: \textit{[Screen, Time, Date, Battery]}.

\end{itemize}


\subsection{Overview of Features}

Our review of the 55 studies revealed inconsistencies in feature construction. Some studies relied solely on statistical features, while others incorporated meaningful features but lacked organized granularity of human behaviour. Time spans were often inconsistently applied or not mentioned at all. This area lacks a clear principle for designing effective features.

After analyzing current mobile sensing studies, we found that effective features for human-centered mobile sensing typically consist of three components: the time span of the sensing data, the metrics used for measurements, and the specific human behaviours being studied.  For instance, the feature  \textit{"Duration of screen time per weeknight"} includes the metric \textit{"Duration of time"}, the atomic behaviour \textit{"screen"} and the time span \textit{"weeknight"}. Table \ref{tab:features} summarizes commonly used time spans and metrics.

\subsection{Model and Performance}
Analysis of 55 mobile sensing studies shows that most research uses similar machine learning models: \textit{Random Forest}, \textit{Gradient Boosting Machine}, \textit{Linear Regression}, \textit{Gaussian Mixture Model}, \textit{Support Vector Machine}, \textit{Naive Bayes}, \textit{K-nearest Neighbour}, and \textit{Logistic Regression}. Some models, like \textit{Random Forest} and \textit{Gradient Boosting}, handle complex relationships well, making them robust for high-dimensional data.
The primary goal of using these models is to evaluate the effectiveness of features in predicting research objectives. However, providing recommendations on model choice is useful. Knowing the approximate performance level for the research objective helps guide researchers' expectations and decisions.

\section{Multi-Granular Human Behaviour Representation}
\label{sec: multigranularity}

Understanding human behaviours deeply is key to effective feature construction and successful mobile sensing studies. Translating sensing signals, smartphone usage, and context descriptors into specific human behaviours (e.g., emotion, alcohol consumption) remains challenging due to the complex nature of human behaviours \cite{gao2022human,granata2015human}. Many studies overlook this aspect, extracting data like statistical features or trajectory data without considering the broader context. This not only wastes time but also fails to cover all facets of human behaviour comprehensively.

\begin{figure}
    \centering
    \includegraphics[width=0.48\textwidth]{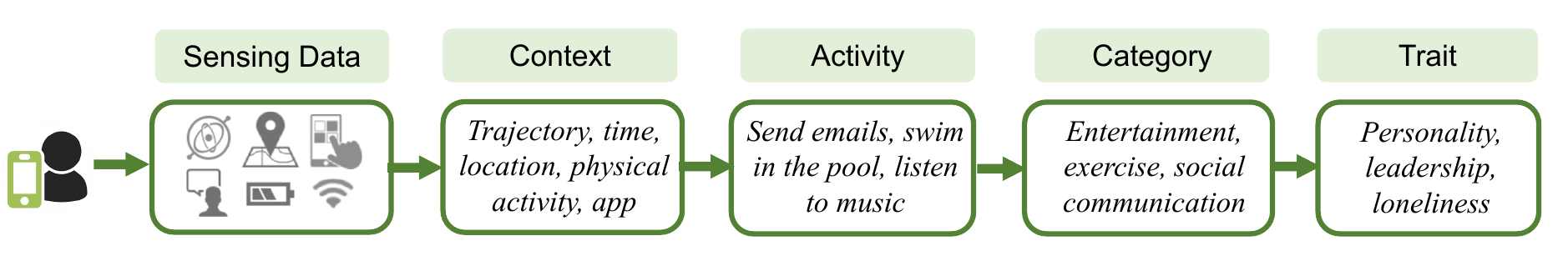}
    \caption{Multi-granular human behaviour representation}
    \label{fig:multi-gran}
\end{figure}

Human behaviour refers to the potential and expressed capacity for physical, mental, and social activity in response to internal and external stimuli throughout life \cite{kagan2020human}. It has been explored by various fields such as psychology, sociology, ethology, and human-centered design. While there are many facets of human behaviour, no single definition or field of study can encapsulate its entirety. For example, behaviour can be decomposed by temporal phases (prenatal life, infancy, childhood, adolescence, adulthood, and old age) \cite{kagan2020human}, reactive modes (reactive and deliberative behaviours) \cite{schmidt2000modelling}, or dimensions (actions, cognition, and emotion). 
To capture human behaviours through smartphones, we propose a multi-granular human behaviour representation mechanism. This mechanism serves as a foundation for constructing meaningful features in mobile sensing research. It comprises four dimensions that encompass human behaviours at varying levels of granularity: contexts, activities, categories, and traits (see Figure \ref{fig:multi-gran}).

\begin{itemize}

    \item \textit{Context:} This level includes information directly inferred or easily calculated from smartphone sensors, such as location/trajectory (GPS), physical activity (Android Activity Recognition API \footnote{Android Activity Recognition API: \url{https://developers.google.com/location-context/activity-recognition}}), time, and screen usage.
        
    \item  \textit{Activity:} This level identifies specific activities or behaviours exhibited by individuals, such as sending emails, swimming, or listening to music.
    
    \item  \textit{Category:} At this level, similar behaviours are grouped based on shared characteristics or attributes, allowing the identification of commonalities and patterns. Categories may include entertainment, exercise, communication, and social activities.

    \item  \textit{Trait:} This level considers enduring characteristics or traits intrinsic to individuals, reflecting their behaviour patterns, such as personality traits, social abilities, leadership, and loneliness.
\end{itemize}

For example, to investigate someone's mood instability (a trait), we can identify relevant categories like stress, happiness, and sadness. Within these categories, activities contributing to mood fluctuations might include work-related tasks, spending time with loved ones, or engaging in hobbies. At the context level, we can examine specific atomic activities like using the smartphone, opening social media apps, texting, and going to bed. By considering these different levels of granularity—from traits to categories to activities and contexts—we can construct a comprehensive representation of human behaviour, enabling a deeper understanding of complex phenomena like mood instability.

\begin{figure}
    \centering
    \includegraphics[width=.5\textwidth]{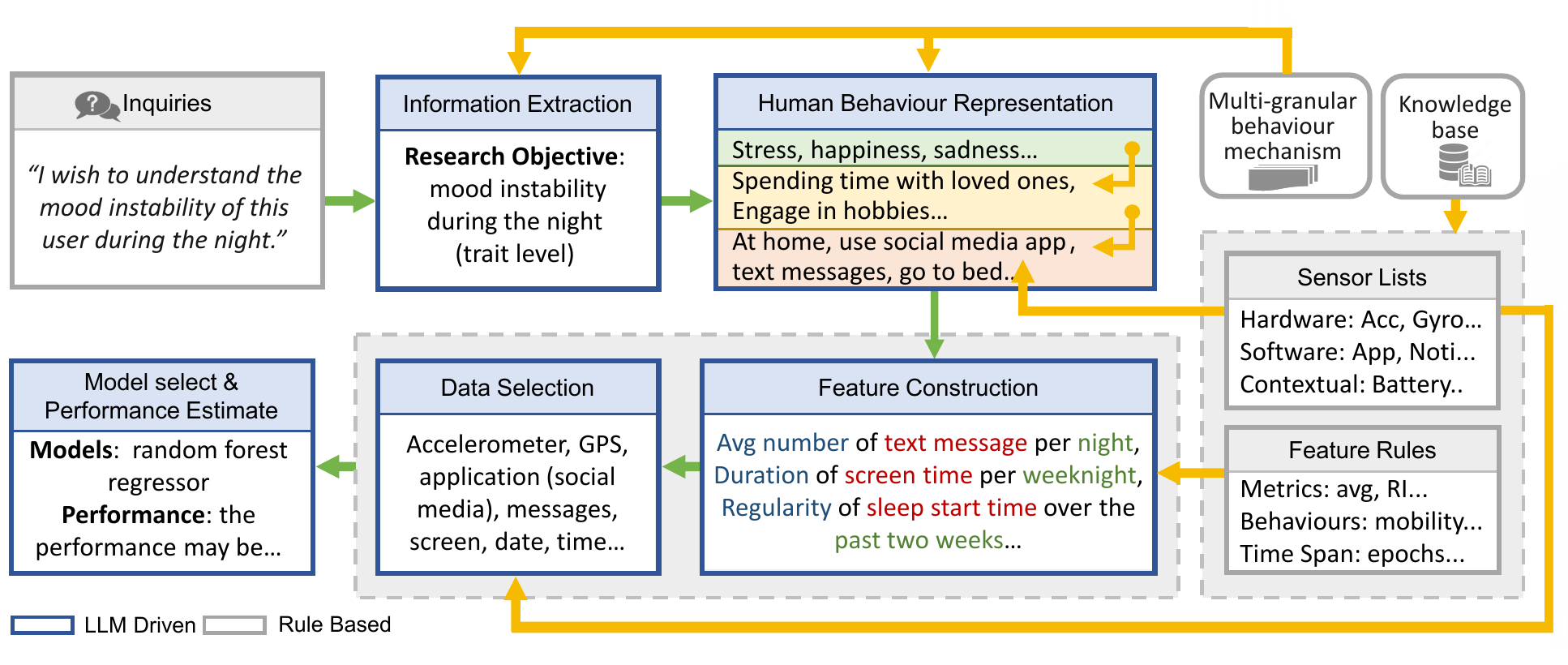}
    \caption{The generation process of mobile sensing strategies involves two main data flows: the user's inquiry in natural language (green arrows) and the designed rules (yellow arrows). These flows merge to produce the final mobile sensing strategies.}
    \label{fig:enter-label}
\end{figure}

\section{Automated Mobile Sensing Framework}
\label{fig: generation process}
\label{sec: strategy}

\subsection{Design Rules for Mobile Sensing Strategies}
\label{subsec: design rules}

To achieve effective mobile sensing strategies, we follow five steps (see Figure \ref{fig: generation process}). The system outputs the strategy based on the user's inquiry after these steps.

\subsubsection{Information Extraction (Step 1)} The user initiates an inquiry, such as \textit{"I wish to understand the mood instability of this user during the night."} The system extracts the research objective, which in this case is \textit{"mood instability during the night"}. Next, the system defines the level of human behaviour (trait, category, activity, or context) based on the multi-granular human behaviour mechanism described in Section \ref{sec: multigranularity}. Since \textit{"mood instability during the night"} is an intrinsic trait affecting behaviour patterns, it is considered at the \textit{trait level}.

\subsubsection{Human Behavior Representation (Step 2)} 
The system extracts multi-granular behaviours based on the objective, moving hierarchically from category to activity and context levels. Context-level behaviours are inferred from smartphone sensing data, using sensor lists from the knowledge base to generate relevant behaviours.

\subsubsection{Feature Construction  (Step 3)} 
The system constructs comprehensive features for modeling the research objective. Effective features consist of the time span of the data, measurement metrics, and specific behaviours. Using context-level behaviours from Step 2, the system selects appropriate metrics and time spans. For instance, a feature for \textit{"mood instability during the night"} could be the \textit{"regularity of sleep start time over the past two weeks"}.

\subsubsection{Data Selection  (Step 4)} 
The system determines the data to compute the features from Step 3. The sensing data source is selected from the identified sensors in Section \ref{subsec: data source}, including hardware, software sensor data, and contextual information. For example, the feature \textit{"regularity of sleep start time"} requires time and sleep data, identified using sensors like the accelerometer and gyroscope.

\subsubsection{Model and its Estimated Performance  (Step 5)}
The system suggests a machine learning model based on the research objective, selected data sources, and constructed features. Good features enhance performance, while limited data sources may hinder accuracy. The system estimates performance using natural language and provides reasoning. For example, modeling psychological traits may result in lower performance due to their complexity and variability.

\subsection{Prompt Structure} 
Our prompt structure, inspired by \cite{wang2023enabling}, consists of three main components: (1) \textbf{Prefix}. A clear and concise introduction outlining the prompt's purpose and design rules, as detailed in Section  \ref{subsec: design rules}, provding a high-level overview and sets the context for the examples that follow.
(2) \textbf{Examples}. 
Each example is divided into three parts:
a) \textit{Inquiry}.  Presents a natural language inquiry to enhance understanding, such as, \textit{"I wish to understand the mood instability of this user during the night."}
b) \textit{Reasoning}. Explains the reasoning behind each design decision with step-by-step justifications, following the design rules.
c) \textit{Mobile Sensing Strategy}.
Outlines the chosen strategies, specifying data to be collected and features to be constructed.
(3) \textbf{User Input}. Users provide their own inquiry related to their objective, expressed in natural language.

\begin{figure}
    \centering
    \includegraphics[width=0.49\textwidth]{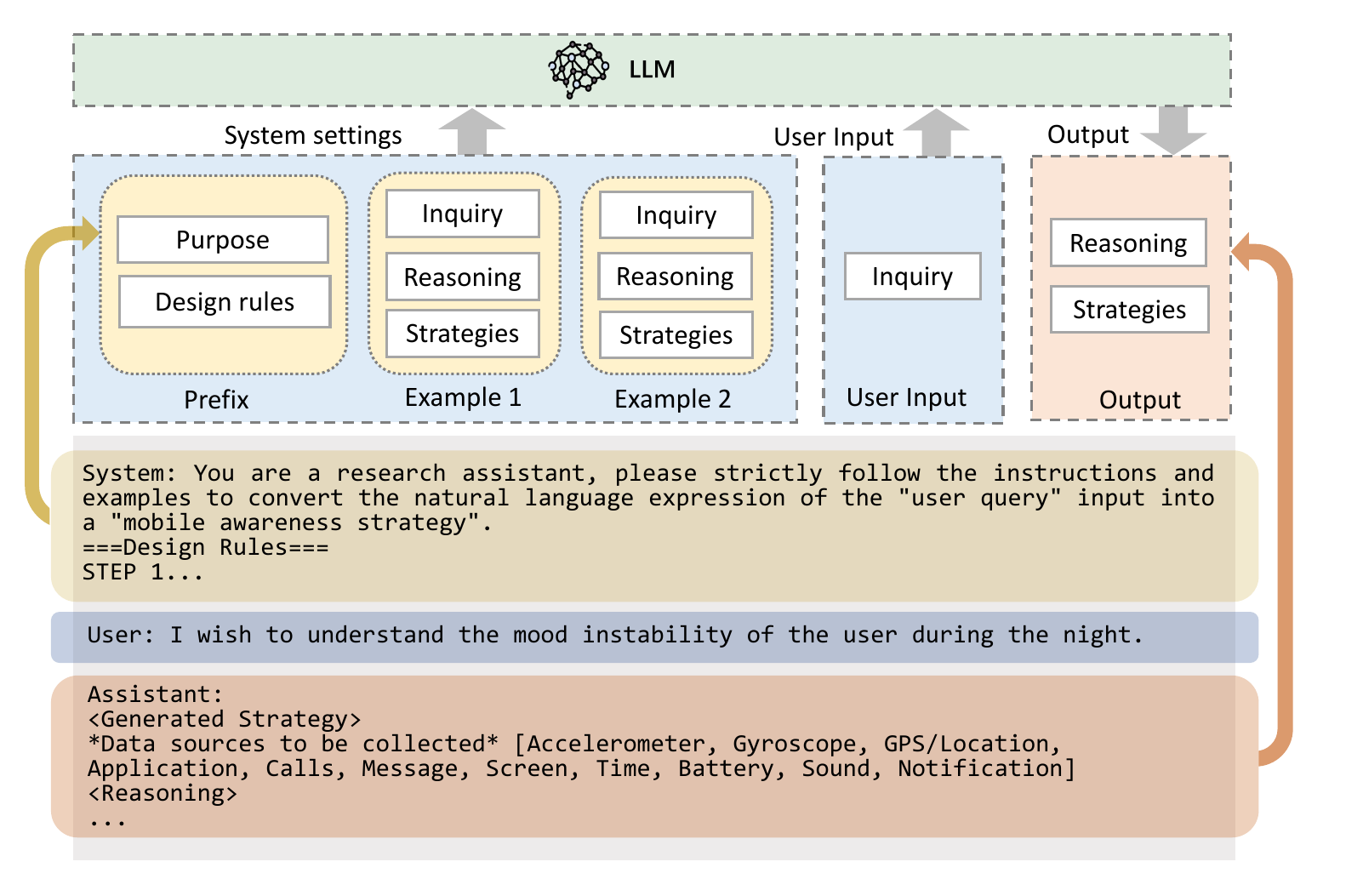}
    \caption{An example illustrating the proposed prompt structure}
    \label{fig:code}
\end{figure}

This approach provides a comprehensive framework that guides LLMs in reasoning and formulating mobile sensing strategies based on the given inquiries. The number of examples can be adjusted according to user requirements. Figure \ref{fig:code} shows an example of our prompt structure. 

\section{Evaluation}
\label{sec: evalution}

In our experiment, we used GPT-3.5-turbo for its robust capabilities in generating coherent, context-rich responses to complex prompts. Users initiated interactions with the model using a prefixed indicator "INPUT". For example, a researcher wishing to model mood instability could type: \textit{INPUT: I wish to understand the mood instability of the user during the night}.

\subsection{Expert Evaluators}

Unlike typical user studies that rely on easily recruited ordinary participants, our research targets experts in the field of mobile sensing to provide valuable insights into human behaviours. We enlisted 8 experts with significant experience in modeling human behaviours using mobile sensing technologies, averaging 4.25 years of research experience. Although the number of evaluators is limited, their profound understanding of mobile sensing techniques provides substantial insights into the system.

\subsection{Procedure}

We conducted two evaluation studies: a comparative study and a usability study. The comparative study evaluated the effectiveness of the automated mobile sensing strategy against existing strategies, using the \textit{Blind Comparison} method \cite{goodwin2016research}. In the usability study, experts typed any inquiry they wanted and then completed a survey and participated in an interview.

\subsubsection{Comparative Study} 


For the comparative study, we selected two highly cited mobile sensing tasks (over 100 citations). We extracted research objectives, selected data sources, and constructed features based on existing descriptions, then applied our sensing strategy generation system. To ensure fairness, we did not include selected models and performance metrics. We maintained consistent data and feature descriptions, excluding sensor details. The primary difference between the existing and automated strategies was the sensing data and features used.

Experts were presented with both existing and auto-generated strategies in a randomized order to avoid \textit{Order Effects} \cite{goodwin2016research}. They compared the strategies, assessing effectiveness, interpretability, relevance, and completeness on a 5-point Likert scale from 1 (very negative) to 5 (very positive). This assessment was conducted through semi-structured interviews, repeated for each expert until both tasks were completed.

\subsubsection{Usability Study} 

In this usability study, experts independently used the automatic sensing strategies system. They typed inquiries into the system to understand various human behaviours through smartphones. The system generated strategies with a step-by-step reasoning process, including data sources to be collected and features to be constructed. We used an adapted NASA-TLX \cite{hoonakker2011measuring} evaluation method to assess the generated strategies, excluding questions on temporal demand or effort as the system required minimal waiting time. Participants rated the following on a 5-point Likert scale, with 1 being the most negative and 5 the most positive:
(1) Mental demand: \textit{How mentally demanding was the task? }
(2) Physical demand: \textit{How physically demanding was the task?}
(3) Performance: \textit{How successful were you in accomplishing what you planned to do?}

Participants also evaluated their overall experience on a 5-point Likert scale, with 1 indicating 'not at all' and 5 indicating 'very much':
(1) Satisfaction: \textit{How satisfied are you with the automated generated strategy?}
(2) Enhanced Understanding: \textit{Does the automated strategy enhance your understanding of the research objective?}
(3) Ease of use: \textit{How easy was the system to use?}
(4) Willingness to reuse: \textit{How likely are you to use this assistant again in the future?}

A concluding interview was conducted to gain deeper insights into the experts' thoughts on the automated mobile sensing strategies, including the system's effectiveness, its impact on their research process, and suggestions for improvements.

\subsection{Result and Discussion}

\begin{figure}
    \subfigure[ {Task: Study A} \label{subfig: age}]{\includegraphics[width=0.23\textwidth]{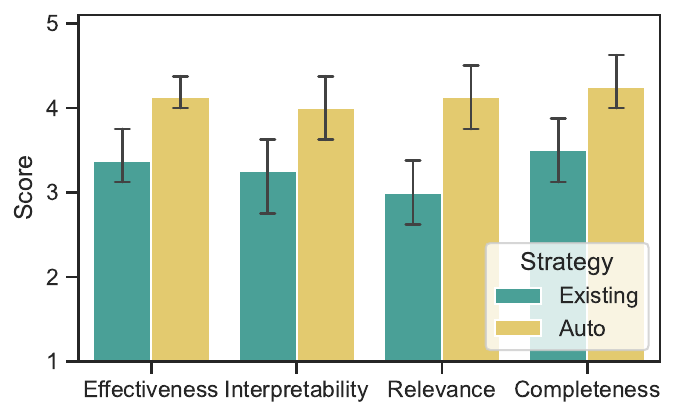}}
	\hspace{0cm}
    \subfigure[ {Task: Study B}\label{subfig: venn}]{\includegraphics[width=0.23\textwidth]{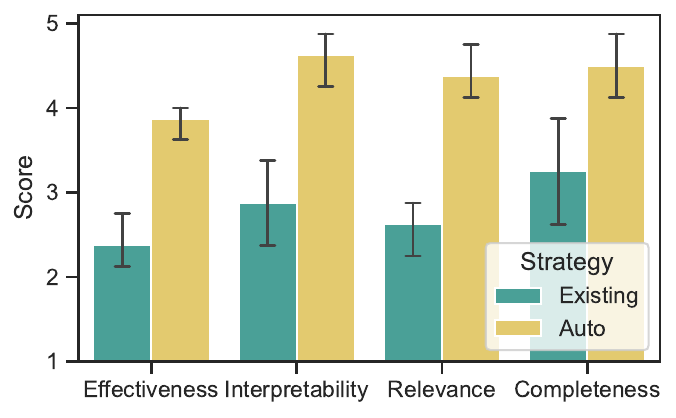}}
    \caption{The evaluation results for both studies from experts}
    \label{fig: comparative evluation}
\end{figure}

\subsubsection{Comparative Performance Analysis}

We compared automated and existing strategies in two studies: predicting \textit{Brain Functional Connectivity} (Study A) and understanding \textit{Compound Emotion} (Study B). Figure \ref{fig: comparative evluation} shows the results for effectiveness, interpretability, relevance, and completeness based on expert opinions. The automated strategy ('Auto') consistently outperformed the existing strategy ('Existing') in all dimensions: 
Effectiveness: Auto 4.0 (STD = 0.37) vs. Existing 2.88 (STD = 0.72).
Interpretability: Auto 4.31 (STD = 0.60) vs. Existing 3.06 (STD = 0.77).
Relevance: Auto 4.25 (STD = 0.58) vs. Existing 2.81 (STD = 0.54).
Completeness: Auto 4.375 (STD = 0.5) vs. Existing 3.375 (STD = 0.81).


Overall, the automated strategies outperformed the existing strategies in all dimensions. However, Performance varied between studies due to different research objectives. Study B's reliance on basic statistical features (e.g., "Longitude, Altitude, Latitude of GPS") was less relevant compared to our proposed features (e.g., "Distance travelled per day/weeknight").

Two experts raised concerns about the feasibility of computing the proposed features, as they are more intricate than low-level statistical features. However, most features can still be computed using mature algorithms. Three experts found the proposed features insightful and beneficial for understanding user behaviour. For example, Expert 2 noted, \textit{"I was pleasantly surprised to find that application data were used in the automated strategy. Obviously, it would be useful for understanding user brain function connectivity"}.


\subsubsection{Usability Analysis}

\begin{figure}
    \centering
    \includegraphics[width=0.45\textwidth]{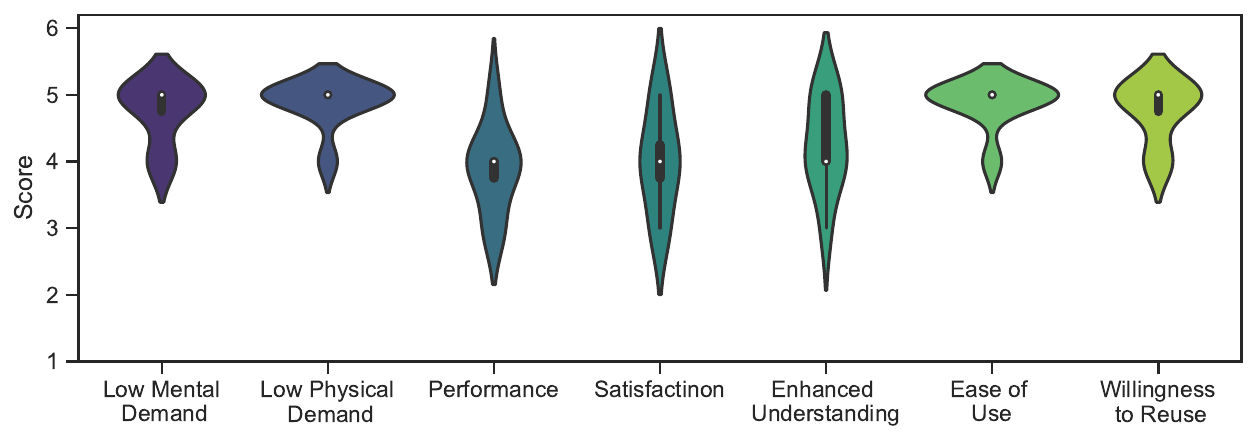}
    \caption{Ratings for the automated mobile sensing strategy from experts}
    \label{fig:nasa}
\end{figure}

Experts tested the system independently and evaluated based on seven dimensions: \textit{Mental Demand}, \textit{Physical Demand}, \textit{Performance}, \textit{Satisfaction}, \textit{Enhanced Understanding}, \textit{Ease of Use}, and \textit{Willingness to Reuse}. As there are no existing automated mobile sensing strategies for comparison, experts rated the system directly on these dimensions. The usability ratings (Figure \ref{fig:nasa}) showed average values above 3 for all dimensions, indicating good performance. Mental and physical demands were low, and experts expressed a strong willingness to use the system for future research.


Experts may propose varying research objectives. While some behaviours are easier to infer (e.g., smartphone addiction), others are more challenging (e.g., heart attack). Despite this, 5 out of 8 experts found the generated strategies meaningful and expressed a desire to use the system for designing their own experiments. Three experts found the proposed features inspiring and valuable. For instance, Expert 5 remarked, \textit{"I was pleasantly surprised to see that application data was incorporated into the automated strategy. Including participants' usage of grooming software would undoubtedly make the experiment more comprehensive"}.

However, there was one case where an expert felt the system's performance was less satisfactory. This dissatisfaction arose from their research objective, which focused on suggesting changes or improvements in behaviour rather than understanding, modeling, or predicting behaviours. Users should ensure that their research objective aligns with understanding human behaviours through mobile sensing. In another scenario, the strategy suggested the feature \textit{"The number of positive/negative messages sent per day"}, which raised two primary concerns: potential violation of privacy rights and ambiguity in distinguishing positive from negative messages. While some generated sensors/features may be valuable, their real-world applicability could be constrained. Further discussion can be found in Section \ref{sec: limiations_implications}.

\section{Implications and Limitations}
\label{sec: limiations_implications}

This research proposes an automatic generation system for mobile sensing strategies to understand human behaviour. For researchers, it reduces the burden of designing strategies, offers effective feature suggestions, and aids decision-making based on estimated performance. The system can adapt to different research objectives, providing tailored suggestions and experimental designs. For individuals, it enhances self-awareness by offering an objective method to understand themselves through passive sensing data, potentially improving well-being and quality of life.

However, this study has limitations. Firstly, not all devices have the same sensors, and availability varies. For example, some devices lack barometers or thermometers, and iOS devices generally have more constraints than Android devices.
Secondly, the study did not cover parameter tuning and data cleaning, focusing instead on data source selection, feature construction, model building, and performance estimation. Tools like AutoML can manage parameter tuning. Thirdly, the research centers on designing automatic mobile sensing strategies without computing or implementing features. The main contribution is strategy design, saving researchers time and reducing the burden of data selection and feature construction, paving the way for future studies.
Lastly, privacy is a concern when collecting data. Although this study does not involve actual data collection, future researchers should implement necessary privacy protection measures. Data processed for automated human behaviour computation would be strictly protected and processed on the user's device, minimizing privacy concerns.

\section{Conclusion}
\label{sec: conclusion}

This paper introduces an automated mobile sensing strategy generation system that allows users to input inquiries related to understanding human behaviours through smartphones.  This automation reduces the burden on researchers and provides new insights for mobile sensing strategy design. Future work will explore automatic feature computation to develop intelligent systems that understand human behaviour, ultimately assisting individuals in gaining self-awareness and enhancing well-being.

\section{Acknowledgments}
This work is supported by the Natural Science Foundation of China (Grant No. 62302252) and the China Postdoctoral Science Foundation (Grant No. 2023M731949).


\bibliographystyle{ACM-Reference-Format}
\bibliography{Nan}


\begin{thebibliography}{42}


\ifx \showCODEN    \undefined \def \showCODEN     #1{\unskip}     \fi
\ifx \showDOI      \undefined \def \showDOI       #1{#1}\fi
\ifx \showISBNx    \undefined \def \showISBNx     #1{\unskip}     \fi
\ifx \showISBNxiii \undefined \def \showISBNxiii  #1{\unskip}     \fi
\ifx \showISSN     \undefined \def \showISSN      #1{\unskip}     \fi
\ifx \showLCCN     \undefined \def \showLCCN      #1{\unskip}     \fi
\ifx \shownote     \undefined \def \shownote      #1{#1}          \fi
\ifx \showarticletitle \undefined \def \showarticletitle #1{#1}   \fi
\ifx \showURL      \undefined \def \showURL       {\relax}        \fi
\providecommand\bibfield[2]{#2}
\providecommand\bibinfo[2]{#2}
\providecommand\natexlab[1]{#1}
\providecommand\showeprint[2][]{arXiv:#2}

\bibitem[Adler et~al\mbox{.}(2021)]%
        {adler2021identifying}
\bibfield{author}{\bibinfo{person}{Daniel~A Adler}, \bibinfo{person}{Vincent
  W-S Tseng}, \bibinfo{person}{Gengmo Qi}, \bibinfo{person}{Joseph Scarpa},
  \bibinfo{person}{Srijan Sen}, {and} \bibinfo{person}{Tanzeem Choudhury}.}
  \bibinfo{year}{2021}\natexlab{}.
\newblock \showarticletitle{Identifying mobile sensing indicators of
  stress-resilience}.
\newblock \bibinfo{journal}{\emph{Proceedings of the ACM on interactive,
  mobile, wearable and ubiquitous technologies}} \bibinfo{volume}{5},
  \bibinfo{number}{2} (\bibinfo{year}{2021}), \bibinfo{pages}{1--32}.
\newblock


\bibitem[Bardram(2020)]%
        {bardram2020carp}
\bibfield{author}{\bibinfo{person}{Jakob~E Bardram}.}
  \bibinfo{year}{2020}\natexlab{}.
\newblock \showarticletitle{The CARP mobile sensing framework--A
  cross-platform, reactive, programming framework and runtime environment for
  digital phenotyping}.
\newblock \bibinfo{journal}{\emph{arXiv preprint arXiv:2006.11904}}
  (\bibinfo{year}{2020}).
\newblock


\bibitem[Chen et~al\mbox{.}(2023)]%
        {chen2023you}
\bibfield{author}{\bibinfo{person}{Yu-Chun Chen}, \bibinfo{person}{Yu-Jen Lee},
  \bibinfo{person}{Kuei-Chun Kao}, \bibinfo{person}{Jie Tsai},
  \bibinfo{person}{En-Chi Liang}, \bibinfo{person}{Wei-Chen Chiu},
  \bibinfo{person}{Faye Shih}, {and} \bibinfo{person}{Yung-Ju Chang}.}
  \bibinfo{year}{2023}\natexlab{}.
\newblock \showarticletitle{Are You Killing Time? Predicting Smartphone
  Users’ Time-killing Moments via Fusion of Smartphone Sensor Data and
  Screenshots}. In \bibinfo{booktitle}{\emph{Proceedings of the 2023 CHI
  Conference on Human Factors in Computing Systems}}. \bibinfo{pages}{1--19}.
\newblock


\bibitem[Dehkordi et~al\mbox{.}(2020)]%
        {dehkordi2020feature}
\bibfield{author}{\bibinfo{person}{Maryam~Banitalebi Dehkordi},
  \bibinfo{person}{Abolfazl Zaraki}, {and} \bibinfo{person}{Rossitza Setchi}.}
  \bibinfo{year}{2020}\natexlab{}.
\newblock \showarticletitle{Feature extraction and feature selection in
  smartphone-based activity recognition}.
\newblock \bibinfo{journal}{\emph{Procedia Computer Science}}
  \bibinfo{volume}{176} (\bibinfo{year}{2020}), \bibinfo{pages}{2655--2664}.
\newblock


\bibitem[Ferreira et~al\mbox{.}(2015)]%
        {ferreira2015aware}
\bibfield{author}{\bibinfo{person}{Denzil Ferreira}, \bibinfo{person}{Vassilis
  Kostakos}, {and} \bibinfo{person}{Anind~K Dey}.}
  \bibinfo{year}{2015}\natexlab{}.
\newblock \showarticletitle{AWARE: mobile context instrumentation framework}.
\newblock \bibinfo{journal}{\emph{Frontiers in ICT}}  \bibinfo{volume}{2}
  (\bibinfo{year}{2015}), \bibinfo{pages}{6}.
\newblock


\bibitem[Feurer et~al\mbox{.}(2020)]%
        {feurer2020auto}
\bibfield{author}{\bibinfo{person}{Matthias Feurer}, \bibinfo{person}{Katharina
  Eggensperger}, \bibinfo{person}{Stefan Falkner}, \bibinfo{person}{Marius
  Lindauer}, {and} \bibinfo{person}{Frank Hutter}.}
  \bibinfo{year}{2020}\natexlab{}.
\newblock \showarticletitle{Auto-sklearn 2.0: The next generation}.
\newblock \bibinfo{journal}{\emph{arXiv preprint arXiv:2007.04074}}
  \bibinfo{volume}{24} (\bibinfo{year}{2020}).
\newblock


\bibitem[Friedman(2002)]%
        {friedman2002stochastic}
\bibfield{author}{\bibinfo{person}{Jerome~H Friedman}.}
  \bibinfo{year}{2002}\natexlab{}.
\newblock \showarticletitle{Stochastic gradient boosting}.
\newblock \bibinfo{journal}{\emph{Computational statistics \& data analysis}}
  \bibinfo{volume}{38}, \bibinfo{number}{4} (\bibinfo{year}{2002}),
  \bibinfo{pages}{367--378}.
\newblock


\bibitem[Gao(2022)]%
        {gao2022human}
\bibfield{author}{\bibinfo{person}{Nan Gao}.} \bibinfo{year}{2022}\natexlab{}.
\newblock \emph{\bibinfo{title}{Human behaviour sensing and profiling in the
  wild}}.
\newblock \bibinfo{thesistype}{Ph.\,D. Dissertation}. \bibinfo{school}{Ph. D.
  Dissertation. RMIT University}.
\newblock


\bibitem[Gao et~al\mbox{.}(2020)]%
        {gao2020n}
\bibfield{author}{\bibinfo{person}{Nan Gao}, \bibinfo{person}{Wei Shao},
  \bibinfo{person}{Mohammad~Saiedur Rahaman}, {and} \bibinfo{person}{Flora~D
  Salim}.} \bibinfo{year}{2020}\natexlab{}.
\newblock \showarticletitle{n-gage: Predicting in-class emotional, behavioural
  and cognitive engagement in the wild}.
\newblock \bibinfo{journal}{\emph{Proceedings of the ACM on Interactive,
  Mobile, Wearable and Ubiquitous Technologies}} \bibinfo{volume}{4},
  \bibinfo{number}{3} (\bibinfo{year}{2020}), \bibinfo{pages}{1--26}.
\newblock


\bibitem[Gao et~al\mbox{.}(2019)]%
        {gao2019predicting}
\bibfield{author}{\bibinfo{person}{Nan Gao}, \bibinfo{person}{Wei Shao}, {and}
  \bibinfo{person}{Flora~D Salim}.} \bibinfo{year}{2019}\natexlab{}.
\newblock \showarticletitle{Predicting personality traits from physical
  activity intensity}.
\newblock \bibinfo{journal}{\emph{Computer}} \bibinfo{volume}{52},
  \bibinfo{number}{7} (\bibinfo{year}{2019}), \bibinfo{pages}{47--56}.
\newblock


\bibitem[Goodwin and Goodwin(2016)]%
        {goodwin2016research}
\bibfield{author}{\bibinfo{person}{Kerri~A Goodwin} {and}
  \bibinfo{person}{C~James Goodwin}.} \bibinfo{year}{2016}\natexlab{}.
\newblock \bibinfo{booktitle}{\emph{Research in psychology: Methods and
  design}}.
\newblock \bibinfo{publisher}{John Wiley \& Sons}.
\newblock


\bibitem[Granata et~al\mbox{.}(2015)]%
        {granata2015human}
\bibfield{author}{\bibinfo{person}{Consuelo Granata}, \bibinfo{person}{Aurelien
  Ibanez}, {and} \bibinfo{person}{Philippe Bidaud}.}
  \bibinfo{year}{2015}\natexlab{}.
\newblock \showarticletitle{Human activity-understanding: A multilayer approach
  combining body movements and contextual descriptors analysis}.
\newblock \bibinfo{journal}{\emph{International Journal of Advanced Robotic
  Systems}} \bibinfo{volume}{12}, \bibinfo{number}{7} (\bibinfo{year}{2015}),
  \bibinfo{pages}{89}.
\newblock


\bibitem[Heinisch et~al\mbox{.}(2022)]%
        {heinisch2022investigating}
\bibfield{author}{\bibinfo{person}{Judith~S Heinisch}, \bibinfo{person}{Nan
  Gao}, \bibinfo{person}{Christoph Anderson}, \bibinfo{person}{Shohreh
  Deldari}, \bibinfo{person}{Klaus David}, {and} \bibinfo{person}{Flora
  Salim}.} \bibinfo{year}{2022}\natexlab{}.
\newblock \showarticletitle{Investigating the Effects of Mood \& Usage
  Behaviour on Notification Response Time}.
\newblock \bibinfo{journal}{\emph{arXiv preprint arXiv:2207.03405}}
  (\bibinfo{year}{2022}).
\newblock


\bibitem[Hollmann et~al\mbox{.}(2023)]%
        {hollmann2023gpt}
\bibfield{author}{\bibinfo{person}{Noah Hollmann}, \bibinfo{person}{Samuel
  M{\"u}ller}, {and} \bibinfo{person}{Frank Hutter}.}
  \bibinfo{year}{2023}\natexlab{}.
\newblock \showarticletitle{GPT for Semi-Automated Data Science: Introducing
  CAAFE for Context-Aware Automated Feature Engineering}.
\newblock \bibinfo{journal}{\emph{arXiv preprint arXiv:2305.03403}}
  (\bibinfo{year}{2023}).
\newblock


\bibitem[Hong et~al\mbox{.}(2022)]%
        {hong2022depressive}
\bibfield{author}{\bibinfo{person}{Juyoung Hong}, \bibinfo{person}{Jiwon Kim},
  \bibinfo{person}{Sunmi Kim}, \bibinfo{person}{Jaewon Oh},
  \bibinfo{person}{Deokjong Lee}, \bibinfo{person}{San Lee},
  \bibinfo{person}{Jinsun Uh}, \bibinfo{person}{Juhong Yoon}, {and}
  \bibinfo{person}{Yukyung Choi}.} \bibinfo{year}{2022}\natexlab{}.
\newblock \showarticletitle{Depressive symptoms feature-based machine learning
  approach to predicting depression using smartphone}. In
  \bibinfo{booktitle}{\emph{Healthcare}}, Vol.~\bibinfo{volume}{10}. MDPI,
  \bibinfo{pages}{1189}.
\newblock


\bibitem[Hoonakker et~al\mbox{.}(2011)]%
        {hoonakker2011measuring}
\bibfield{author}{\bibinfo{person}{Peter Hoonakker}, \bibinfo{person}{Pascale
  Carayon}, \bibinfo{person}{Ayse~P Gurses}, \bibinfo{person}{Roger Brown},
  \bibinfo{person}{Adjhaporn Khunlertkit}, \bibinfo{person}{Kerry McGuire},
  {and} \bibinfo{person}{James~M Walker}.} \bibinfo{year}{2011}\natexlab{}.
\newblock \showarticletitle{Measuring workload of ICU nurses with a
  questionnaire survey: the NASA Task Load Index (TLX)}.
\newblock \bibinfo{journal}{\emph{IIE transactions on healthcare systems
  engineering}} \bibinfo{volume}{1}, \bibinfo{number}{2}
  (\bibinfo{year}{2011}), \bibinfo{pages}{131--143}.
\newblock


\bibitem[Hutter et~al\mbox{.}(2019)]%
        {hutter2019automated}
\bibfield{author}{\bibinfo{person}{Frank Hutter}, \bibinfo{person}{Lars
  Kotthoff}, {and} \bibinfo{person}{Joaquin Vanschoren}.}
  \bibinfo{year}{2019}\natexlab{}.
\newblock \bibinfo{booktitle}{\emph{Automated machine learning: methods,
  systems, challenges}}.
\newblock \bibinfo{publisher}{Springer Nature}.
\newblock


\bibitem[Imambi et~al\mbox{.}(2021)]%
        {imambi2021pytorch}
\bibfield{author}{\bibinfo{person}{Sagar Imambi}, \bibinfo{person}{Kolla~Bhanu
  Prakash}, {and} \bibinfo{person}{GR Kanagachidambaresan}.}
  \bibinfo{year}{2021}\natexlab{}.
\newblock \showarticletitle{PyTorch}.
\newblock \bibinfo{journal}{\emph{Programming with TensorFlow: Solution for
  Edge Computing Applications}} (\bibinfo{year}{2021}),
  \bibinfo{pages}{87--104}.
\newblock


\bibitem[Kagan et~al\mbox{.}(2020)]%
        {kagan2020human}
\bibfield{author}{\bibinfo{person}{Jerome Kagan}, \bibinfo{person}{Marc~H
  Bornstein}, {and} \bibinfo{person}{Richard~M Lerner}.}
  \bibinfo{year}{2020}\natexlab{}.
\newblock \bibinfo{title}{human behaviour. Encyclopedia Britannica}.
\newblock
\newblock


\bibitem[Katevas et~al\mbox{.}(2014)]%
        {katevas2014poster}
\bibfield{author}{\bibinfo{person}{Kleomenis Katevas}, \bibinfo{person}{Hamed
  Haddadi}, {and} \bibinfo{person}{Laurissa Tokarchuk}.}
  \bibinfo{year}{2014}\natexlab{}.
\newblock \showarticletitle{Poster: Sensingkit: A multi-platform mobile sensing
  framework for large-scale experiments}. In
  \bibinfo{booktitle}{\emph{Proceedings of the 20th annual international
  conference on Mobile computing and networking}}. \bibinfo{pages}{375--378}.
\newblock


\bibitem[Kotthoff et~al\mbox{.}(2019)]%
        {kotthoff2019auto}
\bibfield{author}{\bibinfo{person}{Lars Kotthoff}, \bibinfo{person}{Chris
  Thornton}, \bibinfo{person}{Holger~H Hoos}, \bibinfo{person}{Frank Hutter},
  {and} \bibinfo{person}{Kevin Leyton-Brown}.} \bibinfo{year}{2019}\natexlab{}.
\newblock \showarticletitle{Auto-WEKA: Automatic model selection and
  hyperparameter optimization in WEKA}.
\newblock \bibinfo{journal}{\emph{Automated machine learning: methods, systems,
  challenges}} (\bibinfo{year}{2019}), \bibinfo{pages}{81--95}.
\newblock


\bibitem[Kuhn and Johnson(2019)]%
        {kuhn2019feature}
\bibfield{author}{\bibinfo{person}{Max Kuhn} {and} \bibinfo{person}{Kjell
  Johnson}.} \bibinfo{year}{2019}\natexlab{}.
\newblock \bibinfo{booktitle}{\emph{Feature engineering and selection: A
  practical approach for predictive models}}.
\newblock \bibinfo{publisher}{Chapman and Hall/CRC}.
\newblock


\bibitem[Laport-L{\'o}pez et~al\mbox{.}(2020)]%
        {laport2020review}
\bibfield{author}{\bibinfo{person}{Francisco Laport-L{\'o}pez},
  \bibinfo{person}{Emilio Serrano}, \bibinfo{person}{Javier Bajo}, {and}
  \bibinfo{person}{Andrew~T Campbell}.} \bibinfo{year}{2020}\natexlab{}.
\newblock \showarticletitle{A review of mobile sensing systems, applications,
  and opportunities}.
\newblock \bibinfo{journal}{\emph{Knowledge and Information Systems}}
  \bibinfo{volume}{62}, \bibinfo{number}{1} (\bibinfo{year}{2020}),
  \bibinfo{pages}{145--174}.
\newblock


\bibitem[Lee(2014)]%
        {lee2014s}
\bibfield{author}{\bibinfo{person}{Victor~R Lee}.}
  \bibinfo{year}{2014}\natexlab{}.
\newblock \showarticletitle{What's happening in the" Quantified Self"
  movement?}
\newblock \bibinfo{journal}{\emph{ICLS 2014 proceedings}}
  (\bibinfo{year}{2014}), \bibinfo{pages}{1032}.
\newblock


\bibitem[Meegahapola et~al\mbox{.}(2023)]%
        {meegahapola2023generalization}
\bibfield{author}{\bibinfo{person}{Lakmal Meegahapola},
  \bibinfo{person}{William Droz}, \bibinfo{person}{Peter Kun},
  \bibinfo{person}{Amalia De~G{\"o}tzen}, \bibinfo{person}{Chaitanya Nutakki},
  \bibinfo{person}{Shyam Diwakar}, \bibinfo{person}{Salvador~Ruiz Correa},
  \bibinfo{person}{Donglei Song}, \bibinfo{person}{Hao Xu},
  \bibinfo{person}{Miriam Bidoglia}, {et~al\mbox{.}}}
  \bibinfo{year}{2023}\natexlab{}.
\newblock \showarticletitle{Generalization and Personalization of Mobile
  Sensing-Based Mood Inference Models: An Analysis of College Students in Eight
  Countries}.
\newblock \bibinfo{journal}{\emph{Proceedings of the ACM on Interactive,
  Mobile, Wearable and Ubiquitous Technologies}} \bibinfo{volume}{6},
  \bibinfo{number}{4} (\bibinfo{year}{2023}), \bibinfo{pages}{1--32}.
\newblock


\bibitem[Meegahapola et~al\mbox{.}(2021)]%
        {meegahapola2021examining}
\bibfield{author}{\bibinfo{person}{Lakmal Meegahapola},
  \bibinfo{person}{Florian Labhart}, \bibinfo{person}{Thanh-Trung Phan}, {and}
  \bibinfo{person}{Daniel Gatica-Perez}.} \bibinfo{year}{2021}\natexlab{}.
\newblock \showarticletitle{Examining the social context of alcohol drinking in
  young adults with smartphone sensing}.
\newblock \bibinfo{journal}{\emph{Proceedings of the ACM on Interactive,
  Mobile, Wearable and Ubiquitous Technologies}} \bibinfo{volume}{5},
  \bibinfo{number}{3} (\bibinfo{year}{2021}), \bibinfo{pages}{1--26}.
\newblock


\bibitem[Nepal et~al\mbox{.}(2020)]%
        {nepal2020detecting}
\bibfield{author}{\bibinfo{person}{Subigya Nepal}, \bibinfo{person}{Shayan
  Mirjafari}, \bibinfo{person}{Gonzalo~J Martinez}, \bibinfo{person}{Pino
  Audia}, \bibinfo{person}{Aaron Striegel}, {and} \bibinfo{person}{Andrew~T
  Campbell}.} \bibinfo{year}{2020}\natexlab{}.
\newblock \showarticletitle{Detecting job promotion in information workers
  using mobile sensing}.
\newblock \bibinfo{journal}{\emph{Proceedings of the ACM on Interactive,
  Mobile, Wearable and Ubiquitous Technologies}} \bibinfo{volume}{4},
  \bibinfo{number}{3} (\bibinfo{year}{2020}), \bibinfo{pages}{1--28}.
\newblock


\bibitem[Nepal et~al\mbox{.}(2022)]%
        {nepal2022covid}
\bibfield{author}{\bibinfo{person}{Subigya Nepal}, \bibinfo{person}{Weichen
  Wang}, \bibinfo{person}{Vlado Vojdanovski}, \bibinfo{person}{Jeremy~F
  Huckins}, \bibinfo{person}{Alex Dasilva}, \bibinfo{person}{Meghan Meyer},
  {and} \bibinfo{person}{Andrew Campbell}.} \bibinfo{year}{2022}\natexlab{}.
\newblock \showarticletitle{COVID student study: A year in the life of college
  students during the COVID-19 pandemic through the lens of mobile phone
  sensing}. In \bibinfo{booktitle}{\emph{Proceedings of the 2022 CHI Conference
  on Human Factors in Computing Systems}}. \bibinfo{pages}{1--19}.
\newblock


\bibitem[Obuchi et~al\mbox{.}(2020)]%
        {obuchi2020predicting}
\bibfield{author}{\bibinfo{person}{Mikio Obuchi}, \bibinfo{person}{Jeremy~F
  Huckins}, \bibinfo{person}{Weichen Wang}, \bibinfo{person}{Alex Dasilva},
  \bibinfo{person}{Courtney Rogers}, \bibinfo{person}{Eilis Murphy},
  \bibinfo{person}{Elin Hedlund}, \bibinfo{person}{Paul Holtzheimer},
  \bibinfo{person}{Shayan Mirjafari}, {and} \bibinfo{person}{Andrew Campbell}.}
  \bibinfo{year}{2020}\natexlab{}.
\newblock \showarticletitle{Predicting brain functional connectivity using
  mobile sensing}.
\newblock \bibinfo{journal}{\emph{Proceedings of the ACM on interactive,
  mobile, wearable and ubiquitous technologies}} \bibinfo{volume}{4},
  \bibinfo{number}{1} (\bibinfo{year}{2020}), \bibinfo{pages}{1--22}.
\newblock


\bibitem[Rashid et~al\mbox{.}(2020)]%
        {rashid2020predicting}
\bibfield{author}{\bibinfo{person}{Haroon Rashid}, \bibinfo{person}{Sanjana
  Mendu}, \bibinfo{person}{Katharine~E Daniel}, \bibinfo{person}{Miranda~L
  Beltzer}, \bibinfo{person}{Bethany~A Teachman}, \bibinfo{person}{Mehdi
  Boukhechba}, {and} \bibinfo{person}{Laura~E Barnes}.}
  \bibinfo{year}{2020}\natexlab{}.
\newblock \showarticletitle{Predicting subjective measures of social anxiety
  from sparsely collected mobile sensor data}.
\newblock \bibinfo{journal}{\emph{Proceedings of the ACM on Interactive,
  Mobile, Wearable and Ubiquitous Technologies}} \bibinfo{volume}{4},
  \bibinfo{number}{3} (\bibinfo{year}{2020}), \bibinfo{pages}{1--24}.
\newblock


\bibitem[Rish et~al\mbox{.}(2001)]%
        {rish2001empirical}
\bibfield{author}{\bibinfo{person}{Irina Rish} {et~al\mbox{.}}}
  \bibinfo{year}{2001}\natexlab{}.
\newblock \showarticletitle{An empirical study of the naive Bayes classifier}.
  In \bibinfo{booktitle}{\emph{IJCAI 2001 workshop on empirical methods in
  artificial intelligence}}, Vol.~\bibinfo{volume}{3}. \bibinfo{pages}{41--46}.
\newblock


\bibitem[Schmidt(2000)]%
        {schmidt2000modelling}
\bibfield{author}{\bibinfo{person}{Bernard Schmidt}.}
  \bibinfo{year}{2000}\natexlab{}.
\newblock \bibinfo{booktitle}{\emph{The modelling of human behaviour}}.
  Vol.~\bibinfo{volume}{132}.
\newblock \bibinfo{publisher}{Society for Computer Simulation International}.
\newblock


\bibitem[Segal(2004)]%
        {segal2004machine}
\bibfield{author}{\bibinfo{person}{Mark~R Segal}.}
  \bibinfo{year}{2004}\natexlab{}.
\newblock \showarticletitle{Machine learning benchmarks and random forest
  regression}.
\newblock  (\bibinfo{year}{2004}).
\newblock


\bibitem[Tornede et~al\mbox{.}(2023)]%
        {tornede2023automl}
\bibfield{author}{\bibinfo{person}{Alexander Tornede}, \bibinfo{person}{Difan
  Deng}, \bibinfo{person}{Theresa Eimer}, \bibinfo{person}{Joseph Giovanelli},
  \bibinfo{person}{Aditya Mohan}, \bibinfo{person}{Tim Ruhkopf},
  \bibinfo{person}{Sarah Segel}, \bibinfo{person}{Daphne Theodorakopoulos},
  \bibinfo{person}{Tanja Tornede}, \bibinfo{person}{Henning Wachsmuth},
  {et~al\mbox{.}}} \bibinfo{year}{2023}\natexlab{}.
\newblock \showarticletitle{AutoML in the Age of Large Language Models: Current
  Challenges, Future Opportunities and Risks}.
\newblock \bibinfo{journal}{\emph{arXiv preprint arXiv:2306.08107}}
  (\bibinfo{year}{2023}).
\newblock


\bibitem[van Berkel et~al\mbox{.}(2023)]%
        {van2023awarelight}
\bibfield{author}{\bibinfo{person}{Niels van Berkel}, \bibinfo{person}{Simon
  D’Alfonso}, \bibinfo{person}{Rio Kurnia~Susanto}, \bibinfo{person}{Denzil
  Ferreira}, {and} \bibinfo{person}{Vassilis Kostakos}.}
  \bibinfo{year}{2023}\natexlab{}.
\newblock \showarticletitle{AWARE-Light: a smartphone tool for experience
  sampling and digital phenotyping}.
\newblock \bibinfo{journal}{\emph{Personal and Ubiquitous Computing}}
  \bibinfo{volume}{27}, \bibinfo{number}{2} (\bibinfo{year}{2023}),
  \bibinfo{pages}{435--445}.
\newblock


\bibitem[Wampfler et~al\mbox{.}(2022)]%
        {wampfler2022affective}
\bibfield{author}{\bibinfo{person}{Rafael Wampfler}, \bibinfo{person}{Severin
  Klingler}, \bibinfo{person}{Barbara Solenthaler}, \bibinfo{person}{Victor~R
  Schinazi}, \bibinfo{person}{Markus Gross}, {and} \bibinfo{person}{Christian
  Holz}.} \bibinfo{year}{2022}\natexlab{}.
\newblock \showarticletitle{Affective state prediction from smartphone touch
  and sensor data in the wild}. In \bibinfo{booktitle}{\emph{Proceedings of the
  2022 CHI Conference on Human Factors in Computing Systems}}.
  \bibinfo{pages}{1--14}.
\newblock


\bibitem[Wang et~al\mbox{.}(2023)]%
        {wang2023enabling}
\bibfield{author}{\bibinfo{person}{Bryan Wang}, \bibinfo{person}{Gang Li},
  {and} \bibinfo{person}{Yang Li}.} \bibinfo{year}{2023}\natexlab{}.
\newblock \showarticletitle{Enabling conversational interaction with mobile ui
  using large language models}. In \bibinfo{booktitle}{\emph{Proceedings of the
  2023 CHI Conference on Human Factors in Computing Systems}}.
  \bibinfo{pages}{1--17}.
\newblock


\bibitem[Wang et~al\mbox{.}(2015)]%
        {wang2015smartgpa}
\bibfield{author}{\bibinfo{person}{Rui Wang}, \bibinfo{person}{Gabriella
  Harari}, \bibinfo{person}{Peilin Hao}, \bibinfo{person}{Xia Zhou}, {and}
  \bibinfo{person}{Andrew~T Campbell}.} \bibinfo{year}{2015}\natexlab{}.
\newblock \showarticletitle{SmartGPA: how smartphones can assess and predict
  academic performance of college students}. In
  \bibinfo{booktitle}{\emph{Proceedings of the 2015 ACM international joint
  conference on pervasive and ubiquitous computing}}.
  \bibinfo{pages}{295--306}.
\newblock


\bibitem[Wang et~al\mbox{.}(2017)]%
        {wang2017predicting}
\bibfield{author}{\bibinfo{person}{Rui Wang}, \bibinfo{person}{Weichen Wang},
  \bibinfo{person}{Min~SH Aung}, \bibinfo{person}{Dror Ben-Zeev},
  \bibinfo{person}{Rachel Brian}, \bibinfo{person}{Andrew~T Campbell},
  \bibinfo{person}{Tanzeem Choudhury}, \bibinfo{person}{Marta Hauser},
  \bibinfo{person}{John Kane}, \bibinfo{person}{Emily~A Scherer},
  {et~al\mbox{.}}} \bibinfo{year}{2017}\natexlab{}.
\newblock \showarticletitle{Predicting symptom trajectories of schizophrenia
  using mobile sensing}.
\newblock \bibinfo{journal}{\emph{Proceedings of the ACM on Interactive,
  Mobile, Wearable and Ubiquitous Technologies}} \bibinfo{volume}{1},
  \bibinfo{number}{3} (\bibinfo{year}{2017}), \bibinfo{pages}{1--24}.
\newblock


\bibitem[Wang et~al\mbox{.}(2020)]%
        {wang2020social}
\bibfield{author}{\bibinfo{person}{Weichen Wang}, \bibinfo{person}{Shayan
  Mirjafari}, \bibinfo{person}{Gabriella Harari}, \bibinfo{person}{Dror
  Ben-Zeev}, \bibinfo{person}{Rachel Brian}, \bibinfo{person}{Tanzeem
  Choudhury}, \bibinfo{person}{Marta Hauser}, \bibinfo{person}{John Kane},
  \bibinfo{person}{Kizito Masaba}, \bibinfo{person}{Subigya Nepal},
  {et~al\mbox{.}}} \bibinfo{year}{2020}\natexlab{}.
\newblock \showarticletitle{Social sensing: assessing social functioning of
  patients living with schizophrenia using mobile phone sensing}. In
  \bibinfo{booktitle}{\emph{Proceedings of the 2020 CHI conference on human
  factors in computing systems}}. \bibinfo{pages}{1--15}.
\newblock


\bibitem[Wang et~al\mbox{.}(2022)]%
        {wang2022first}
\bibfield{author}{\bibinfo{person}{Weichen Wang}, \bibinfo{person}{Subigya
  Nepal}, \bibinfo{person}{Jeremy~F Huckins}, \bibinfo{person}{Lessley
  Hernandez}, \bibinfo{person}{Vlado Vojdanovski}, \bibinfo{person}{Dante
  Mack}, \bibinfo{person}{Jane Plomp}, \bibinfo{person}{Arvind Pillai},
  \bibinfo{person}{Mikio Obuchi}, \bibinfo{person}{Alex Dasilva},
  {et~al\mbox{.}}} \bibinfo{year}{2022}\natexlab{}.
\newblock \showarticletitle{First-gen lens: Assessing mental health of
  first-generation students across their first year at college using mobile
  sensing}.
\newblock \bibinfo{journal}{\emph{Proceedings of the ACM on interactive,
  mobile, wearable and ubiquitous technologies}} \bibinfo{volume}{6},
  \bibinfo{number}{2} (\bibinfo{year}{2022}), \bibinfo{pages}{1--32}.
\newblock


\bibitem[Y{\"u}r{\"u}ten et~al\mbox{.}(2014)]%
        {yuruten2014predictors}
\bibfield{author}{\bibinfo{person}{Onur Y{\"u}r{\"u}ten},
  \bibinfo{person}{Jiyong Zhang}, {and} \bibinfo{person}{Pearl~HZ Pu}.}
  \bibinfo{year}{2014}\natexlab{}.
\newblock \showarticletitle{Predictors of life satisfaction based on daily
  activities from mobile sensor data}. In \bibinfo{booktitle}{\emph{Proceedings
  of the SIGCHI Conference on Human Factors in Computing Systems}}.
  \bibinfo{pages}{497--500}.
\newblock


\end{thebibliography}



\end{document}